\documentclass[]{article}

\usepackage{color}
\usepackage[usenames,dvipsnames,svgnames,table]{xcolor}
\usepackage{graphicx}

\small

\begin{document}

\twocolumn

\begin{center}
\fboxrule0.02cm
\fboxsep0.4cm
\fcolorbox{blue}{AliceBlue}{\rule[-0.9cm]{0.0cm}{1.8cm}{\parbox{7.8cm}
{ \begin{center}
{\Large\em My Favorite Object}

\vspace{0.2cm}

{\large\bf L1551 IRS 5} 

%\vspace{0.05cm}

%{\large\bf }

\vspace{0.2cm}

{\large\em Malcolm Fridlund}

\vspace{0.5cm}

%\begin{figure}[h!]
\centering
\includegraphics[width=0.21\textwidth]{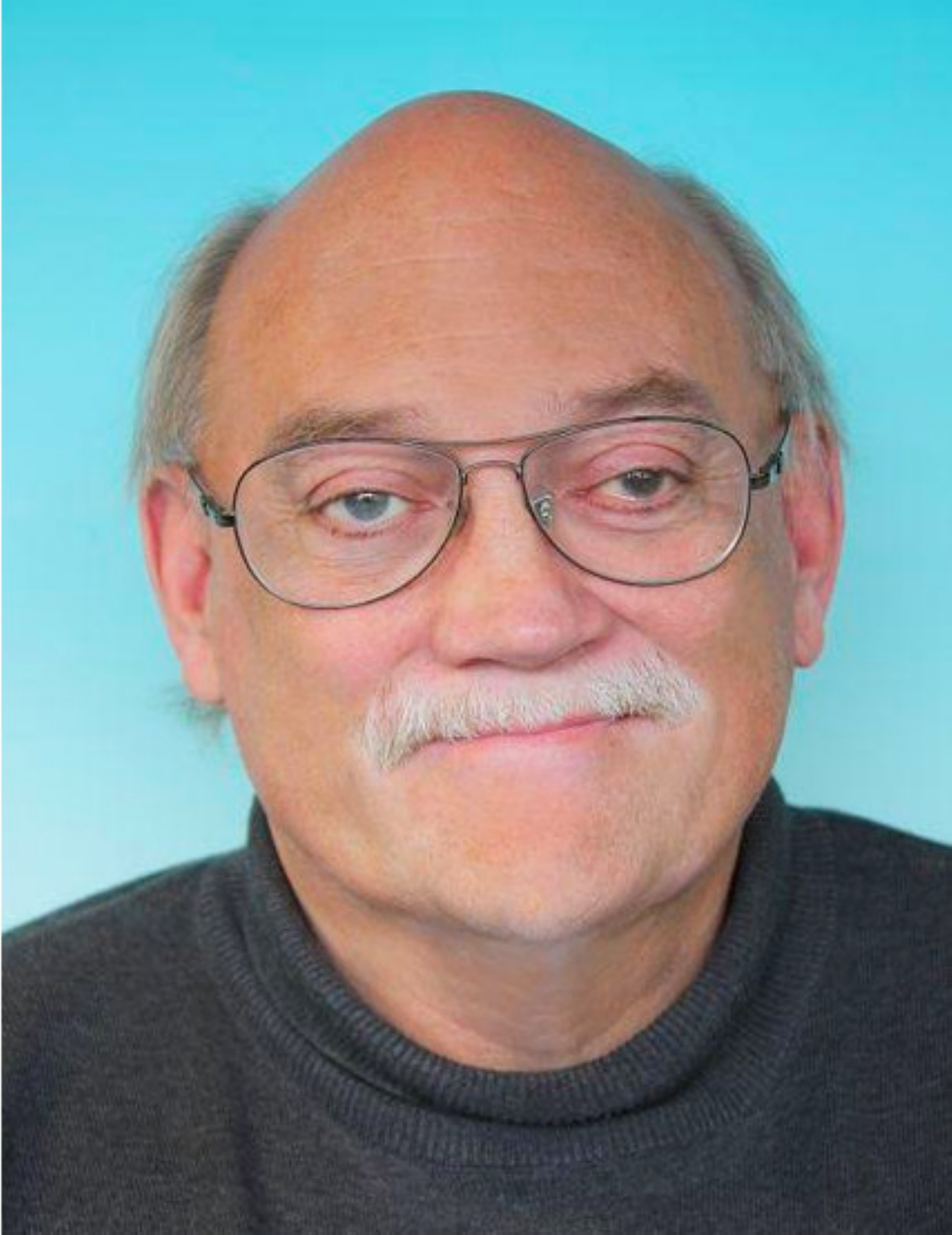}
%\end{figure}
\end{center}
}}}
\end{center}

%\subsection*{header}

{\large\bf Introduction}

The L1551-IRS 5 source and its associated molecular and atomic outflow
have been in focus of front-line research for at least the last 40
years. It was the first object I observed professionally and wrote my first paper on it. Opus No.1
 as it were. I found it so interesting that after my my thesis about its 
molecular outflow I have now published more than
20 papers about this object and I am still fascinated by it.

I find it a very important object to study because of its
proximity and orientation, and its display of basically all tracers of
(low-mass) star formation such as its location at the edge of a dense
molecular cloud and its associated Herbig-Haro objects. It has an
atomic jet and an aligned molecular outflow that displays well
separated blue- and red-shifted outflow lobes. The orientation of the
outflow is thus at a shallow angle with respect to the plane of the
sky making the resolution of small spatial elements possible. Like the
majority of stellar systems, the source IRS 5 is even a binary, each
component of which have a solar-system size protostellar disk (Bieging
\& Cohen, 1985; Rodr\'iguez et al., 1986; Looney et al., 1997;
Rodr\'iguez et al., 1998), or maybe even a triple star system (Lim \&
Takakuwa, 2006). Apparently each of the components have its associated
atomic jet (Fridlund \& Liseau, 1998), and it has been suggested that
there is also individual molecular outflows coming from each component
(Wu et al., 2009). This assembly is wrapped inside a larger disk (of
roughly Kuiper-Edgeworth belt dimensions) which in itself is found
inside a rotating, flattened, molecular gas envelope of Oort cloud
dimensions (Fridlund et al., 2002). The two atomic jets emanate from
IRS5 A and B and can be traced for some thousands of AU (tens of arc
seconds). IRS5 and its associated jet(s) is also one of the first
Herbig-Haro type sources where X-ray emission has been discovered
(Favata et al., 2002). Herbig Haro jets can thus carry enough energy
to create X-ray sources that can be used to further study important
physical parameters of low mass jets and their interaction with the
ambient medium, and even on short timescales (Favata et al., 2006).
 
 All of this together make the L1551 IRS 5 object a marvellous
 laboratory in order to study the early phases of star formation and
 especially how the angular momentum of a collapsing cloudlet is the
 driving engine for a large number of phenomena accompanying this
 process. While stellar formation may be understood in principle, much of
 the details remain unknown, partly because it takes place hidden
 behind large amounts of extinction, requiring the development of more
 and more sophisticated techniques in the IR. 
 It would be impossible to cover everything that has been studied and
 written about this fascinating object here. There are much more than
 100 refereed papers written primarily about different aspects of
 L1551 IRS5 and its associated objects between 1979 and now. Therefore
 such an endeavour will have to await a proper review article (which
 appear to be well motivated), and the present article will have to be
 more of a personal account, sampling certain aspects of L1551 IRS 5.

\begin{figure*}[htb]
\begin{center}
\includegraphics[width=0.8\textwidth]{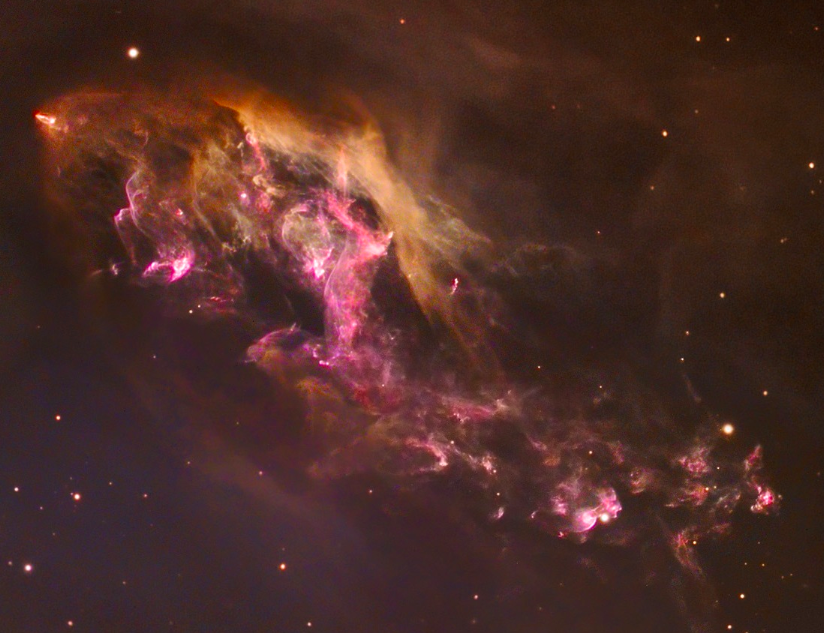}
% OBS: FIGURES MUST BE PS OR EPS FILES
\caption{The blueshifted outflow from L1551 IRS5 (upper left) has
  burst through the cloud surface and reveals intricate shocks as
  Herbig-Haro objects.  Deep H$\alpha$ and [SII] images obtained by Bo
  Reipurth at the 8m Subaru telescope; color composite by Robert
  Gendler. }
\label{....}
\end{center}
\end{figure*}

\vspace{0.5cm}
 
 {\large\bf The molecular cloud L1551}

 This molecular cloud was noted more than 60 years ago.  Herbig
 obtained the first photographic plates of the region in 1951
 (Cudworth \& Herbig, 1979), and Sharpless (1959) listed the brightest
 optical nebulosity visible against the cloud as number 239 (the
 T-Tauri nebulosity is No 238) in his catalogue of HII regions.The
 molecular cloud L1551 itself is about 30 arc minutes across (Lynds,
 1965), and over the years a number of other important young objects
 such as L1551 NE, HL \& XZ Tau, the HH30 disk and its associated jet,
 IR sources and eventually several molecular outflows and Herbig-Haro
 objects have become associated with this cloud. In this article I
 will restrict myself almost exclusively to L1551 IRS 5 and its
 associated objects. Under the name of S239, Knapp et al. (1976)
 carried out very early CO and $^{13}$CO observations with the 11m
 NRAO telescope at Kitt Peak. They found large velocity gradients
 which they interpreted as the signature of infall of the whole cloud
 (about 100 $M_{\odot}$ of molecular material). These observations led
 Snell et al. (1980) to make a more detailed study of these molecular
 tracers, using the 4.9m MWO antenna in Texas, and mapping the
 kinematics they discovered the first of many bipolar molecular
 outflows.

 Strom et al. (1974) studied HH objects with the objective of
 acquiring more information about the embedded IR sources that had
 been discovered during several surveys in the late 1960s, and for
 this purpose they included the HH28, HH29 and HH102 objects. 
% near what was going to be known as L1551 IRS 5. 
% Their hypothesis -- only partly right -- were that HH objects were
% reflection nebulae very close to IR sources, and as such would
% provide the spectrum of the embedded objects themselves. This turns
% out to be rarely true but it is indeed correct as what concerns 
 HH102 (ex S239 -- see Mundt et al. 1985) turns out to be a reflection
 nebula, not an HH object (large yellowish nebula to the upper right
 in Fig.1).  Strom et al. (1976) identified a 2.2$\mu$m source, that
 became known as IRS 5. More or less simultaneously, Cudworth \&
 Herbig (1979) had observed two Herbig-Haro objects clearly associated
 with the L1551 cloud, HH28 and HH29, and, comparing their data with
 photographic plates obtained in the beginning of the fifties, found
 that HH28 and HH29 showed significant proper motions. The proper
 motion vectors also appeared to intersect, when plotted backwards
 several hundred years in time, at the position of the IRS 5! Finally,
 again more or less simultaneously, Sandqvist \& Bernes (1980) had
 published observations at 2-mm, 2-cm and 6-cm of H$_2$CO,
 formaldehyde, that could be modelled as an excellent tracer of
 density enhancements in molecular clouds, and found a density knot
 exactly on the position of L1551 IRS 5. This peak in the density
 indicated that we were dealing with a heavily embedded object , which
 made it a very nice target for observations in the far infrared
 (FIR). And Sandqvist \& Bernes were in Stockholm....

\vspace{0.3cm}
 
 {\large\bf Balloon observations from Texas....}

 At the end of 1978 I was just finishing my undergraduate studies in
 Stockholm and was thinking about what to do next when I was
 approached by Lennart Nordh (then at the Stockholm Observatory) who
 wanted to know if I was interested in a job. Three weeks later I had
 moved to Groningen in Holland and was busy acclimatising myself in
 the FIR balloon group there. Fast forward another five months (May
 1979) and I was sitting at the National Center for Atmospheric
 Research (NCAR) balloon base in Palestine, Texas, carrying out my own
 first observation of L1551 IRS 5 with a balloon borne 60cm telescope,
 hanging under a balloon the size of a football field, and operating
 at 42 km altitude. The photometer had filters transmitting between
 72$\mu$m and 196$\mu$m. From these observations we derived a total
 flux from IRS 5 of 25 solar luminosities (Fridlund et al., 1980),
 assuming a distance of 150pc, (L1551 is today considered to be at a
 distance of 140 pc (Kenyon et al. 1994). So the
 conclusion was that it was a low mass pre-main-sequence stellar
 object, associated with the molecular outflow. I remained in Holland
 until 1981 working on a total of 6 balloon flights observing mainly
 compact, embedded HII regions. When I returned to Sweden in early
 1982 I had a lot of material for my theses but my advisor (again
 Lennart Nordh) argued that the launch of the IRAS satellite just
 around the corner could possibly make my thesis obsolete. He
 suggested follow-up observations of 1-5$\mu$m NIR of my sources, as
 well as using the 20m mm-telescope at Onsala, Sweden for CO and
 $^{13}$CO observations with much higher spatial resolution than
 achieved by Snell et al. (1980).

\begin{figure}[htb]
\begin{center}
\includegraphics[width=0.45\textwidth]{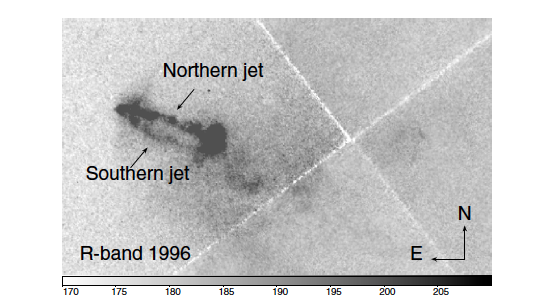}
% OBS: FIGURES MUST BE PS OR EPS FILES
\caption{ R-band HST image of the HH 154 southern and northern jets in 1996. }
\label{....}
\end{center}
\end{figure}

\begin{figure}[htb]
\begin{center}
\includegraphics[width=0.45\textwidth]{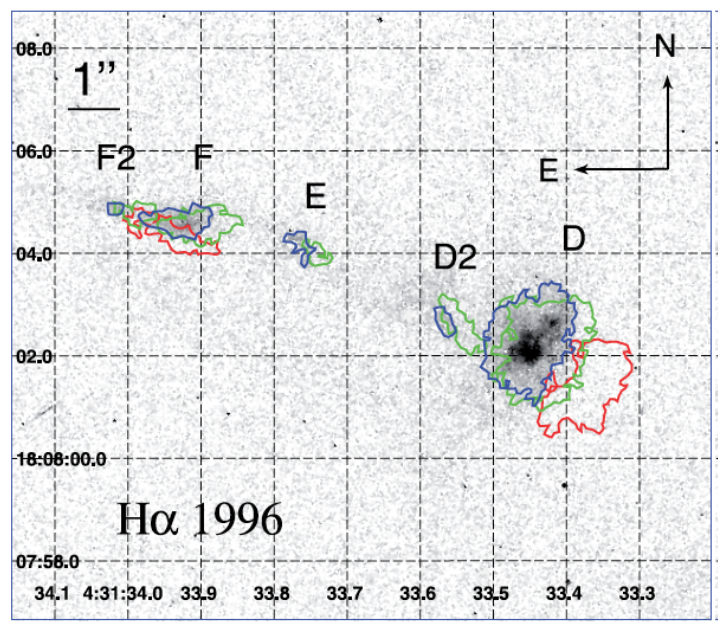}
% OBS: FIGURES MUST BE PS OR EPS FILES
\caption{ HST image of the HH 154 jet in 1996 in the H$\alpha$. The
  superimposed contours define the knots within the jet at different
  epochs: 1996 (blue), 1998 (green), and 2005 (red). }
\label{....}
\end{center}
\end{figure}

%\begin{figure}[htb]
%\begin{center}
%\includegraphics[width=0.5\textwidth]{fig1.ps}
% OBS: FIGURES MUST BE PS OR EPS FILES
%\caption{....}
%\label{....}
%\end{center}
%\end{figure}

\vspace{0.3cm}
 
 {\large\bf ....Groundbased and satellites}

 The NIR observations carried out with the InSb photometer on the ESO
 1m telescope in La Silla indicated that IRS 5 had lost more than 1/2
 a magnitude in the K-band (2.2$\mu$m) with respect to the
 observations carried out by Strom et al. (1976) which could be
 indications that a FU~Orionis type outburst had taken place recently
 (Fridlund et al., 1984b; Fridlund, 1987). The CO observations took
 place under some of the best conditions ever experienced at Onsala
 and I got beautiful data (Fridlund et al., 1984a). It was obvious to
 me that this was what my thesis should really be about and I should
 map the complete outflow at high spatial resolution. As mentioned
 above, the red- and blue-shifted CO outflow lobes are spatially well
 separated and have an angular extent of more than 20 arcminutes on
 the sky. The beam size of the Onsala 20m at the 2.6mm wavelength of
 CO/$^{13}$CO is 30" and I was planning to use a beam spacing of 20"
 so it was a large project under any conditions, and since the
 exceptional conditions experienced during my first run never
 rematerialised it meant that I did not finish the mapping of the
 blue-shifted outflow lobe until 1986. Those results were published
 half a year later as my PhD thesis (Fridlund, 1987). The red-shifted
 outflow lobe was eventually only partially sampled by me. My results
 demonstrated, nevertheless, clearly that the molecular outflow is
 mainly composed of the swept-up gas along the surface of the outflow
 lobe, a result that was confirmed by a more complete study by
 Moriarty-Scheiven et al. in a series of papers (1987a, 1987b, 1988).
 The outflow thus has the structure of evacuated edge-enhanced lobes,
 but it was found around this time that these lobes are filled with
 neutral atomic (HI) gas participating in the outflow (Lizano et al.,
 1988). Parenthetically, in January 1983 The Infrared Astronomical
 Satellite (IRAS) was launched. This was the first-ever space-based
 observatory and during ten months (until the liquid He cooling the
 whole spacecraft boiled off) it mapped essentially the whole sky at
 12, 25, 60 and 100$\mu$m wavelength. Although it produced much better
 quality data than our balloon telescope, not a single one of our
 results were changed except providing smaller error bars. IRAS did,
 however, discover a new embedded FIR source, that became known as
 L1551 NE about 2.5 arcminutes away from IRS 5 and much weaker -- only
 9 $L_{\odot}$ (Emerson et al., 1984). I remembered something when
 reading this paper, and looked up the original data from the balloon
 flight in 1979, and there was L1551 NE, just next to IRS 5 and just
 barely spatially resolved from it -- but with a S/N of only 2.5
 (although seen in independent scans in both filters)! So sometimes
 noise is data (and vice versa) and one has to know which is which.

 In the meantime a lot had happened also in the visual wavelength
 range. Image intensified photographic plates and the first CCD
 detectors had become available. With such detectors, optical studies
 could be carried out in a fraction of the time previously needed when
 using photographic plates alone. In a large study, the nebulosity
 very near ($<$ 10 arcseconds) IRS 5 was interpreted as either a jet
 or an edge-enhanced cavity by Mundt \& Fried (1983), who also found
 that it possessed a very knotted structure. Further studies by Mundt
 et al. (1985) who obtained spectra of the HH 102 found
 characteristics of a reflection nebulosity that displayed FU Orionis
 characteristics (i.e. P Cygni profiles) that they could associate
 with IRS 5. This was consistent with the 1-5$\mu$m observations by
 Fridlund (1984b, 1987). Sarcander at al., (1985) carried out a
 spectroscopic investigation of the IRS5 jet (as well as the HH28,
 HH29 and HH102 nebulosity) and found very high radial velocities in
 the brightest knot in the jet, as well as being able to estimate the
 density of the jet for the first time, and calculate that the jet
 only possessed at most 1\% of the required momentum to drive the
 molecular outflow. Neckel \& Staude (1987) then obtained new imaging
 data and comparing with the old images from Mundt \& Fried in 1983,
 they could detect clear morphological changes in the jet consistent
 with the velocities $> 200~km s^{-1}$~found in the spectroscopy of
 Mundt et al (1985) and the proper motions of Sarcander et al (1985).
 Stocke et al. (1988) also studied the jet spectroscopically
 confirming the high radial velocities and also finding strong
 indications that the visual extinction towards IRS 5 itself could be
 as high as 150 magnitudes or more. Given the geometry of the outflow,
 as well as the presence of the HH objects and their velocity vectors,
 this could immediately be interpreted as having a dense dusty disk
 orbiting the protostar and being viewed edge on (at right angles to
 the major axis of the outflow). Observations in the near IR (Campbell
 et al. 1988) could also be interpreted in this framework.

 Almost all of the Herbig Haro objects are located within the 'blue'
 CO lobe. The most prominent objects are HH28, HH29 and HH102 are also
 found within the cavity walls of the 'blue' lobe. Also the IRS 5 jet,
 which has Herbig Haro characteristics (beginning with Sarcander et
 al. 1985) and which led to it receiving the designation HH154 in the
 compilation of Reipurth 1999).  Essentially only HH 262 (L\'opez et
 al., 1998) can be found within the 'red' lobe. That more objects are
 visible in the 'blue' lobe is not so strange since apparently the
 outflow is emanating out of the cloud here while the 'red' lobe is
 penetrating into the L1551 cloud itself experiencing a progressively
 higher extinction. Liseau et al. (2005) found that the outflow is
 oriented with its major outflow axis most likely at an angle of
 between 45 and 60 degrees w.r.t. the plane of the sky, The velocity
 gradients found in CO (Fridlund et al., 1984a), as well as the radial
 velocity field of both HH29 and that found in the jet (Fridlund \&
 Liseau, 1998; Fridlund et al., 1998), then made it very clear that
 the outflow that is originating at IRS 5 is interacting with the
 ambient medium at the position of the HH objects.  The proper motion
 vectors of HH 28 and HH29 had also been found to intersect at IRS 5
 when projected backwards in time by Cudworth \& Herbig (1979).  By
 observing HH29 (which is unusually bright for an HH object) with the
 International Ultraviolet Explorer (IUE) satellite, Liseau et al.
 (1996) found that the individual knots within HH 29 were varying by
 large factors over times less than 6 weeks (interval between
 observations).  This indicated a very small size scale for the
 interacting elements, and a time scale for the shocking gas to pass
 through ambient knots (which must have a density enhancement of at
 least $10^3$ to produce the UV spectrum), of less than 6 weeks.
 
 It should be
 noted that Devine et al. (1999, 2000) have found that there appear to
 be more outflows somewhat criss-crossing each other in the region and
 those authors argue that some of the HH objects, specifically HH28
 and HH29, could be excited by a flow from another YSO (most probably
 L1551NE). More studies are needed to clarify this. Here we only note
 that both HH28 and HH29 are spatially within the 'blue' lobe of the
 outflow originating at IRS 5, as defined by the molecular studies,
 and this outflow is without doubt the most energetic in the region.
 The proper motion studies by Cudworth \& Herbig (1979) also support
 this scenario (see below).
 
 Because of the advent of a new detector developed by ESA for the
 Hubble Space Telescope another opportunity arose. I was now (since
 1989) working for ESA and one of my tasks was to take a development
 of this HST detector and test it on the ground. The detector was a
 very sensitive photon counting detector, which equipped with very
 narrow band filters could be used to study the emission lines in
 nebulae quantitatively. Bringing this instrument to the Nordic
 Optical telescope in La Palma, Spain, we began by observing HH29 in a
 number of shock diagnostic emission lines (Fridlund et al., 1993) and
 continued with observations of the IRS 5 jet, now using a CCD camera
 instead since they had become more sensitive (Fridlund \& Liseau,
 1994), where we confirmed and extended the work of Neckel \& Staude
 (1987). The results were so promising it led us to propose a
 multi-cycle program on the HST with its high spatial resolution.  We
 followed up our observations of HH29 with a program further
 clarifying the physical parameters such as densities and velocities
 using the NTT/EMMI echelle spectrograph at ESO, La Silla. We found
 densities in the individual knots of HH 29 of 10$^3$ cm$^{-3}$ -
 10$^4$ cm$^{-3}$ within an 'inter-clump' medium with a density of
 about 300 cm$^{-3}$, as well as radial velocities of up to 200 km
 s$^{-1}$.

\begin{figure}[htb]
\begin{center}
\includegraphics[width=0.5\textwidth]{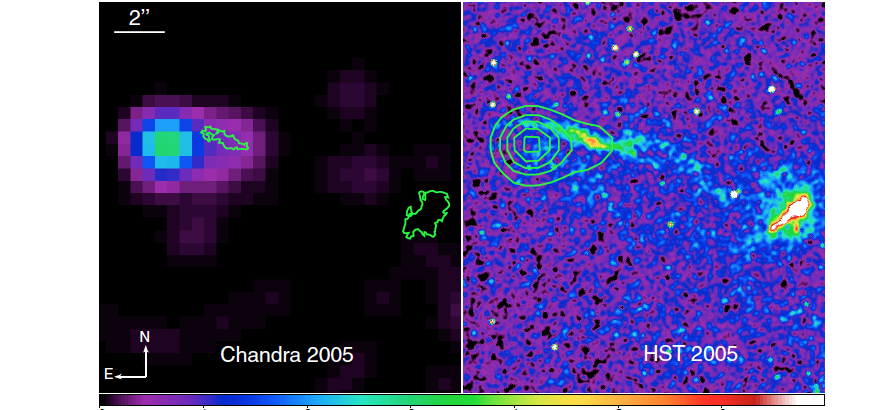}
% OBS: FIGURES MUST BE PS OR EPS FILES
\caption{ X-ray (left panel) and H$\alpha$ (right panel) emission as
  detected with Chandra and Hubble, respectively. The contours of the
  optical knots and of the X-ray emission are superimposed on the
  X-ray and optical images, respectively.  }
\label{....}
\end{center}
\end{figure}

 The binary of IRS 5 was noted early (Bieging \& Cohen, 1985;
 Rodr\'iguez et al., 1986) in radio observations penetrating the
 extinction, although it was debated for a few years whether one was
 observing a toroid shaped gas disk or the components in a binary. It
 was interferometric observations that finally could decide that this
 was a case of two stars being formed within the same envelope (Looney
 et al., 1997; Rodr\'iguez et al., 1998) and eventually tracing the
 optical jets disappearing behind the obscuration, each towards one of
 the two components (Rodr\'iguez et al., 2003a) and finally observing
 the proper motions of the components and determine their total mass
 to be 1.2 $M_{\odot}$ (Rodr\'iguez et al., 2003b). These data were
 later used by Liseau et al. (2005) to model the physical parameters
 of the system further, arriving at stellar masses of 0.8
 $M_{\odot}$~and 0.3$M_{\odot}$~respectively.

\vspace{0.3cm}
 
 {\large\bf Into deepest space}

 Our first set of HST WFPC2 observations were carried out in cycles 5
 \& 6, the results of which can be found in a series of papers. The
 first results were published in Fridlund \& Liseau (1998). We
 confirmed the binarity of the jet and could connect it to the
 binarity of IRS 5 itself, as well as identify the two separate
 velocity systems belonging to the two jets. We further derived the
 structure of the working surface of the jet (HH154), resolving
 completely the Mach disk from the bow shock. Comparing with shock
 models, and having the radial velocities (from observations with the
 Nordic Optical Telescope) and an estimate for the proper motions of
 the working surface (from the results quoted above) it was possible
 to derive the electron density of the jet and we could (again)
 conclude the jet could not be the driver for the large scale
 molecular outflow since {\em a)} it is lacking at least two orders of
 magnitude in momentum and {\em b)} its dynamic age is 3 orders of
 magnitude too small thus confirming the results of Sarcander et al
 (1985). In Fridlund et al. (2005) we presented the complete analysis
 of the HST data from these cycles. We found that the highest 'true'
 velocities within the jet were between 500 and 600~km s$^{-1}$~ as
 well as mapping out the appearance, disappearance and apparent
 movements of different features within the jet. Specifically we found
 that the Mach disk of the working surface is moving downstream at a
 transverse velocity of 180~km s$^{-1}$.  In 2002 we observed L1551
 with ESA's X-ray satellite XMM. I was expecting, based on the strong
 shock characteristics we had found in HH29, that it was most likely
 that we would find any X-rays there. It was, however, within the jet
 itself that we discovered an X-ray source (Favata et al., 2002) Later
 observations (Bally et al., 2003 and Favata et al., 2006)
 demonstrated that the position of the X-ray source is about 0.8 arc seconds
 away from IRS 5 itself (about 100 AU), just where the jets emerge
 from behind the optical extinction. Representative of shock
 temperatures of 4MK, it appears just where we observe 'true'
 velocities of 500 to 600~s$^{-1}$.

\begin{figure}[htb]
\begin{center}
\includegraphics[width=0.45\textwidth]{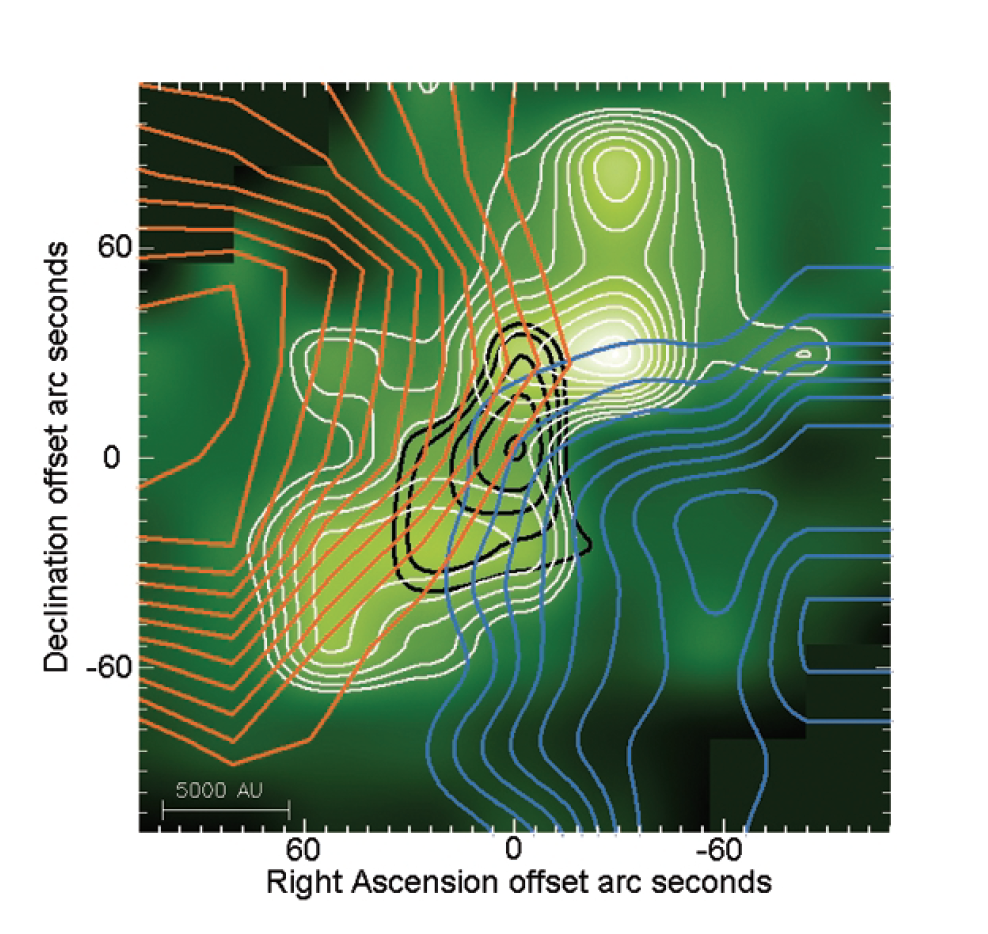}
% OBS: FIGURES MUST BE PS OR EPS FILES
\caption{ CH$_3$OH J = 2-1 line map (green image with white contours)
  superimposed with contours of the HCN J = 1-0 (Onsala data shown as
  black contours) from White et al., 2006, and red- and blue shifted CO J = 1-0 outflows
  (shown as red and blue contour lines, respectively from Fridlund et al.,
  2002). The bottom left bar shows the size scale at 140 pc distance.  }
\label{....}
\end{center}
\end{figure}

\vspace{0.3cm}
 
 {\large\bf Molecules again}

 We observed the inner 3 arc minutes by 3 arc minutes around IRS 5 in
 a number of molecules, HCO$^+$, H$^{13}$CO$^+$, $^{12}$CO and
 $^{13}$CO, all in the J=1-0 transition, using the Onsala 20m
 telescope and acquiring very high signal to noise observations
 (Fridlund et al., 2002). By using all these isotopomers we could
 determine the total mass of the gas, its kinematics and the general
 structure in a region about the size of the Oort cloud in the Solar
 System. By tracing the self reversal in the HCO$^+$, we believe we
 have defined the mid plane of the disk. The H$^{13}$CO$^+$, on the
 other hand is optically thin or near so and have been used to
 calculate the mass of the disk and its surrounding envelope, which
 turns out to be about 2.5 $M_{\odot}$ (excluding the protostars).
 Kinematically most of this molecular material appears to be rotating,
 an observation strongly supporting that of Kaifu et al (1984) which
 was based on observations of CS with the Nobeyama 45m telescope. We
 also find that the molecular outflow has a very high mass loss rate,
 ~ 10$^{-5}$$M_{\odot}~yr^{-1}$ confirming the result of Fridlund \& Knee
 (1993).

 We also searched for gas phase methanol (CH$_3$OH) and HCN, in the direction of L1551 IRS
 5 expecting to see it only directly at the position of the protostar
 itself, since it should freeze out under the conditions of the
 disk/envelope. Surprisingly we discovered it essentially all along
 the rotating flattened envelope (White et al., 2006, see Fig. 5). The only
 observable heat source that could provide the UV or harder photons is
 the jet, and modelling the geometry of the situation and knowing the
 flux of HH154, we found that radiation from the jet could
 sufficiently heat the surface of the disk tens of thousands of AU
 away from the jet base.
 
 I would like to finish with some general thoughts about what we see
 in the area around L1551. Many sets of data covering wavelengths from
 the X-ray region to cm radio wavelengths are in agreement that some
 very energetic processes are taking place here at different spatial
 scales. One may assume that the energy required for these processes
 is coming from the accretion process, but the gravitational energy
 released from the collapse of the part of the cloud forming IRS 5 is
 only part of it. The rotational energy from the large flattened
 envelope has to be removed as the gas is slowly accreting towards
 the centre of the forming star system. There is a tremendous amount
 of angular momentum that need to be transferred into the outflow. The
 mechanism doing this is still unknown. There may also be a large
 amount of magnetic energy 'frozen' into the rotating gas envelope,
 and where the diffusion time is significantly longer than the star
 formation process. The molecular outflow seems to have 2-3 orders of
 magnitude more momentum than the optically visible atomic jet could
 deliver to it so it can not -- at least not currently -- be providing
 all the input of energy needed into the outflow. There are three
 components of the outflow that need to be coupled together. The
 visible jet with velocities of up to more than 500 ~km s$^{-1}$, the
 HI flow filling the swept-up cavity with velocities around 150 ~km
 s$^{-1}$ and the molecular outflow towards the edges with velocities
 of around 10-50 ~km s$^{-1}$.  All of these issues, which probably
 also occur in other star forming regions during the outflow phase,
 are, in my opinion, elements where the continuing study of L1551 IRS
 5 may bring further clarity.

 This source has continued to surprise me for a very long time now and
 I am certain that new observations, especially with ALMA, and the
 continued monitoring in the optical and X-rays will continue to do so.

\footnotesize

{\bf References:}

Bally, J., et al., 2003, ApJ, 584, 843\\
Bieging, J.H., Cohen, M., 1985, ApJ, 289, L5\\
Bonito, R., et al, 2008, A\&A, 484, 389\\
Campbell, B., et al., 1988, AJ, 95, 1173\\
Cudworth, K.M., Herbig, G., 1979, AJ, 84, 548\\
Devine, D., et al.,1999, AJ, 118, 972\\
Devine, D., et al.,  2000, AJ 119, 1872\\
Emerson, J.P., et al., 1984, ApJ, 278, L49\\
Favata, F., et al., 2002, A\&A, 386, 204\\
Favata, F., et al., 2006, A\&A, 450, 17\\
Fridlund, M., et al., 1980, A \&A, 91, L1\\
Fridlund, M., et al., 1984a, A\&A, 137 L17\\
Fridlund et al., 1984b, Stockholm Observatory report No. 26\\
Fridlund, M. 1987, ph.D theses,  University of Stockholm\\
Fridlund, M., et al., 1989, A\&A, 213, 310\\
Fridlund M., Knee, L., 1993, A\&A, 268, 245\\
Fridlund, M., et al., 1993 A\&A, 273, 601\\
Fridlund, M., Liseau, R., 1994, A\&A, 292, 631\\
Fridlund M., Liseau, R., 1998, Ap.J., 499, L75\\
Fridlund, M., et al, 1998, A\&A, 330, 327\\
Fridlund, M., et al., 2002, A\&A, 382, 573\\
Kaifu, N., et al., 1984, A\&A, 134, 7\\
Knapp, G. R. et al., 1976, ApJ, 206, 443\\
Kenyon, S., et al., 1994,  AJ, 108, 1872K\\
Lizano, S., et al., 1988, ApJ, 328, 763\\
Liseau, R, Sandell, G., 1986, ApJ, 304, 459\\
Liseau, R., et al., 1996, A\&A, 306, 255\\ 
Liseau, R., et al., 2005, Ap.J., 619, 959\\
Looney, L.W., et al., 1997, ApJ, 484, L157\\
L\'opez, R., et al., 1998, AJ, 116, 845\\
Lynds, B.T., 1965, ApJS, 12, 163\\ 
Moriarty-Schieven, G., et al., 1987a, ApJ, 317, L95\\
Moriarty-Schieven, G., et al., 1987b, ApJ, 319, 742\\
Moriarty-Schieven, G., 1988, phD theses, University of Massachusets\\
Moriarty-Schieven, G., Snell, R., 1988, ApJ, 332, 364\\
Moriarty-Schieven, G., Wannier, P., 1991, ApJ, 373, L23\\
Mundt, R., Fried 1983, ApJ, 274, L83\\
Mundt, R., et al., 1985, ApJ, 297, L41\\
Neckel, Th., Staude, J., 1987, ApJ, 322, 27\\
Reipurth, B., 1999, , http:casa.colorado.edu/hhcat\\
Rodr\'iguez, L.F., et al., 1986, ApJ, 301, L25\\
Rodr\'iguez, L.F., et al., 1998, Nature, 395, 355\\
Rodr\'iguez, L.F., et al., 2003a, ApJ, 586, L137\\
Rodr\'iguez, L.F., et al., 2003b, ApJ, 583,330\\
Sandqvist, Aa., Bernes, C., 1980, A\&A, 89, 187\\
Sarcander, M., et al., 1985, ApJ, 288, L51\\
Sharpless, S., 1959, ApJS, 4, 257 \\
Snell, R., et al. 1980, ApJ, 239, 17L\\
Stocke, J.T., et al., 1988, ApJS, 68, 229\\
Strom, S.E., et al., 1974, ApJ, 191, 111\\
Strom, K., et al., 1976, AJ, 81, 320\\
White, G.J. et al., 2006, ApJ, 651, L41\\
Wu, P-F., et al., 2009, ApJ, 698, 184\\
% Argiroffi, C., et al. 2012, ApJ, 752, 100\\
% Byrne, P.B. 1986, Irish AJ, 17, 294\\

\normalsize

\end{document}